\documentclass[prc,preprint,tightenlines,showpacs,nofootinbib]{revtex4}

\usepackage{graphicx}  
\usepackage{bm}  
\usepackage{amsmath}
\usepackage{epsf}
\newcommand{\beq}{\begin{equation}}
\newcommand{\eeq}{\end{equation}}
\newcommand{\bea}{\begin{eqnarray}}
\newcommand{\eea}{\end{eqnarray}}

\newcommand{\simgt}{\stackrel{>}{{}_\sim}}

\begin{document}
\title{Universality in the triton charge form factor}
\author{L. Platter}\email{lplatter@phy.ohiou.edu}
\affiliation{Helmholtz-Institut f\"ur Strahlen- und Kernphysik (Theorie),
Universit\"at Bonn, Nu\ss allee 14-16, D-53115 Bonn, Germany}
\affiliation{Department of Physics and Astronomy, Ohio University,
Athens, OH 45701, USA}
\author{H.-W. Hammer}\email{hammer@itkp.uni-bonn.de}
\affiliation{Helmholtz-Institut f\"ur Strahlen- und Kernphysik (Theorie),
Universit\"at Bonn, Nu\ss allee 14-16, D-53115 Bonn, Germany}
\date{\today}
\begin{abstract}
We consider the three-nucleon system within an effective theory with
contact interactions at leading order in the large scattering
length. We calculate the charge form factor of the triton
at low momentum transfer and extract the triton charge radius.
At this order, no two-body currents contribute and the calculation
can be performed in the impulse approximation. We also comment on the 
power counting for higher orders.
The requirement of a three-body force for renormalization of the
three-nucleon system
explains the previously observed correlation between the triton 
binding energy and charge radius for different model potentials.
\end{abstract}
\pacs{21.45.+v, 21.10.Dr, 21.10.Ft}
\maketitle
\section{Introduction}
\label{sec:intro}

There has been a considerable interest in physical systems with
large scattering length recently.
This interest was stimulated by the effective
field theory program in nuclear physics \cite{Beane:2000fx,Bedaque:2002mn}
and by the experimental realization of Feshbach resonances with trapped 
atoms, leading to tunable interactions in atomic systems
\cite{Inouye98,Courteille98,Roberts98}.
The scattering of two particles with short-range interactions
at sufficiently low energy is determined by their S-wave scattering length
$a$. If $a$ is much larger than the typical 
low-energy length scale $\ell$, which is given by the range of
the interaction, the system shows universal
properties. The simplest example is the existence of a shallow 
two-body bound state if $a$ is large and positive, but there are many more, 
including the effects of a limit cycle \cite{Bedaque:1998kg,Bedaque:1998km}
and the Efimov effect \cite{Efimov71,Efimov79} in the three-body system.
For a recent review of the universal properties of $N$-body systems
with short-range interactions up to $N=4$, see Ref.~\cite{Braaten:2004rn}.

The best known example of a nuclear system with a large scattering
length is the two-nucleon ($NN$) system. There are two independent S-wave
scattering lengths that govern the low-energy scattering
of nucleons. The scattering lengths
$a_s= -23.5$ fm and $a_t= 5.42$ fm describe $NN$ scattering in the
spin-singlet ($^1S_0$) and spin-triplet ($^3S_1$) channels, respectively.
Both scattering lengths are significantly larger than the natural
low-energy length scale $\ell\sim 1/m_\pi \approx 1.4$ fm, while
the effective ranges are of the same order as $\ell$.\footnote{
In the following, we will simply use $a$ as a generic symbol for
$a_s$ and $a_t$ if no distinction is required.}
As a consequence, the description of few-nucleon systems in an 
expansion in $\ell/|a|$ is useful. It has successfully been applied 
to various two-, three-, and four-nucleon observables 
(See Refs.~\cite{Beane:2000fx,Bedaque:2002mn,Hammer:2005bp,
Griesshammer:2005ga,Platter:2004zs} and references therein).

The most well-known example of a
universal feature of the three-nucleon system is the Phillips
line \cite{Phillips68}. If the predictions of different nucleon-nucleon
potentials for the triton binding energy $B_t$ and the spin-doublet
neutron-deuteron scattering length  $a_{nd}^{(1/2)}$ are plotted
against each other, they fall close to a line.
This correlation between $B_t$ and $a_{nd}^{(1/2)}$
can not be understood in conventional potential
models. However, it immediately follows from universality if the
large $NN$ scattering lengths are exploited within an expansion in
$\ell/|a|$ \cite{EfiTk85,Bedaque:1999ve}.
If corrections of order $\ell/|a|$ are neglected, all low-energy
3-nucleon observables depend only on the spin-singlet and spin-triplet
scattering lengths $a_s$ and $a_t$ and the three-body parameter $L_3$.
Since the $NN$ potentials reproduce the
scattering phase shifts, they all have the same scattering
lengths. However, the off-shell behavior of the potentials is
not constrained by the phase shifts and in general is different for
each potential. This difference is captured in the three-body parameter $L_3$.
The various potential model calculations must therefore
fall close to a line which is parametrized by the parameter $L_3$.
A similar universal feature of the four-nucleon system is
the Tjon line: an approximately linear correlation between the
triton binding energy $B_t$ and the binding energy of the $\alpha$-particle
 $B_\alpha$. This correlation was discovered
by Tjon \cite{Tjo75} using simple separable interactions,
but also holds for modern phenomenological potentials \cite{Nogga:2000uu}.
The origin of this correlation was explained in Ref.~\cite{Platter:2004zs}
from the absence of a four-body force at leading order in the 
pionless effective field theory. As a consequence, the Tjon line is also
parametrized by the parameter $L_3$.
However, the universal properties resulting from the large scattering
length are not restricted to purely hadronic observables.
For example, a correlation between the triton binding energy and 
charge radius was observed by Friar, Gibson,
Chen, and Payne \cite{Friar85}.

This work is another step towards the treatment of external currents in 
the three-nucleon system within the pionless effective theory. 
We calculate the charge form factor of the triton 
at leading order in $\ell/|a|$ and extract the charge radius.
Furthermore, we study the correlation between the triton binding energy and 
charge radius and relate it to the three-body parameter $L_3$ that 
parametrizes the Phillips and Tjon lines. For a recent calculation of
the cross section for the reaction $nd\to\mbox{$^3$H}\gamma$ at 
energies below 200 keV see Ref.~\cite{Sadeghi:2004es}.

\section{Effective Theory}
\label{sec:eft}

Effective theories are a powerful tool to calculate low-energy observables
in a systematic fashion. They are ideally suited to exploit a separation
of scales such as the one between $|a|$ and $\ell$. At sufficiently
low-energies, one can use an effective theory with contact interactions
only. In this paper, we work in a quantum-mechanical framework and
construct an effective low-energy interaction potential.
In the following, we briefly review this construction. For a more detailed
discussion see Ref.~\cite{Platter:2004qn}.

The bare effective potential generated by short-range contact interactions 
can be written down in a momentum expansion. In the S-wave sector of 
the two-nucleon system, it takes the general form
\beq
\langle {\bf k'} | V_{bare} | {\bf k} \rangle =
{\cal P}_s\,\Bigl[\lambda_2^s\,+\lambda_{2,2}^s(k^2+k'^2)/2\Bigr]
  +{\cal P}_t\,\Bigl[\lambda_2^t\,+ \lambda_{2,2}^t(k^2+k'^2)/2\Bigr]
+\ldots\,,
\label{effpot_bare}
\eeq
where the dots indicate momentum dependent terms that are higher
order in $\ell/|a|$ and ${\cal P}_s$ and ${\cal P}_t$ project onto the
$^1S_0$ and $^3S_1$ partial waves, respectively.
Because of Galilean invariance, the interaction can only depend on the
relative momenta of the incoming and outgoing particles
${\bf k}$ and ${\bf k'}$. 
In a momentum cutoff scheme, the potential in Eq.(\ref{effpot_bare})
can be regularized by multiplying with a regulator function,
$\exp[-(k^2+k'^2)^n /\Lambda^{2n}]$, where $n \geq 1$ is an integer. 
The unphysical cutoff parameter $\Lambda$ is arbitrary and strongly
suppresses the contribution of momentum states with $k,k' \simgt \Lambda$.
Expanding the regulator function shows that the coefficients 
$\lambda_{2,2n}^{s,t}$ of
the momentum expansion of the effective theory at order $(k^2+k'^2)^n$
are modified by the regulator. One can either absorb these additional 
contributions in the definition of the couplings $\lambda_{2,2n}^{s,t}$ or 
choose a value of $n$ such that only terms at orders higher than the 
desired accuracy in the momentum expansion appear \cite{Epelbaum:1999dj}.
In the end, all observables will be independent of $\Lambda$ up to
higher order corrections. At leading order, we can choose a Gaussian
regulator with $n=1$ and the regularized potential takes the form
\beq
\langle {\bf k'} | V_{LO}| {\bf k} \rangle =
{\cal P}_s\,\lambda_2^s\, g({\bf k'})g({\bf k}) +{\cal P}_t\,\lambda_2^t\,
g({\bf k'})g({\bf k})~,
\label{effpot}
\eeq
with $g({\bf k})=\exp(-k^2/\Lambda^2)$.

The interactions in Eq.~(\ref{effpot})
are separable and thus, the two-body problem
for each partial wave can be solved analytically. The two-body
t-matrix can be written as
$t_{t,s}(E)=|g\rangle\tau_{t,s}(E)\langle g|$,
where $E$ denotes the energy. The two-body propagator $\tau_{t,s}(E)$ is
given by:
\begin{equation}
\tau_{t,s}(E)=\Bigl[1/\lambda_2^{t,s}-4\pi\int\hbox{d}q\,
  q^2\frac{g(q)^2}{E-q^2}\Bigr]^{-1}~.
\label{eq-tau}
\end{equation}
Here and in the following we use units with $\hbar=m=1$ for convenience.
The coupling constants $\lambda_2^s$ and $\lambda_2^t$ can be fixed by
demanding that the triplet and singlet scattering lengths $a_t$ and
and $a_s$ are reproduced correctly by the corresponding t-matrices.

The properties of the triton are determined by the Faddeev
equations \cite{Faddeev:1960su}.
To leading order in $\ell/|a|$, all internal orbital angular
momenta can be set to zero. Thus, the triton spin $1/2$ is built
up from the spins of the nucleon only.
The triton wave function can be decomposed into Faddeev components.
We can eliminate all but one of the components and obtain
an equation for the remaining component $\psi_1$ :
\beq
\psi_1=G_0 t P \psi_1+G_0 t G_0 t_3 (1+P)\psi_1\,,
\label{eq:fadd}
\eeq
where $G_0$ is the free three-particle propagator and
$P=P_{13}P_{23}+P_{12}P_{23}$ is a permutation operator that
generates the omitted Faddeev components; $P_{ij}$ exchanges
particles $i$ and $j$. The full wave function $\Psi$ can be recovered in
the end from the equation 
\beq
\Psi=(1+P)\,\psi_1+G_0t_3(1+P)\psi_1\,.
\label{eq:fullwf}
\eeq
The auxilliary quantity
$t_3$ is the solution of a Lippmann-Schwinger equation with
a $SU(4)$-symmetric three-body contact interaction
\beq
\langle {\bf u_1}\:{\bf u_2}|V_3|{\bf u'_1}\:{\bf u'_2}\rangle
={\cal P}_a \lambda_3 h(u_1,u_2) h(u'_1,u'_2)
\eeq
only. Here ${\cal P}_a$ denotes the projector on the total antisymmetric
three-body state with total spin $S=1/2$ and total isospin $T=1/2$
as for example given in \cite{Delfino:1984}.
The regulator function 
$h({\bf u_1},{\bf u_2})=\exp(-(u_1^2+{\textstyle \frac{3}{4}}u_2^2)/
\Lambda^2)$ is defined in terms of the familiar Jacobi momenta
of the three-body system.
The three-body coupling $\lambda_3$ must be determined from a three-body
observable. This interaction is required in order to renormalize the
three-body system and achieve independence of the cutoff
parameter $\Lambda$.
Its renormalization group behavior is governed
by a limit cycle \cite{Bedaque:1999ve,Platter:2004zs}.
For large values of $\Lambda$ the running of the coupling
constant is described by
\beq
\lambda_3(\Lambda)=\frac{c}{\Lambda^4} \;\frac{\sin(s_0 \ln(\Lambda/L_3)
  -\arctan(1/s_0))}{\sin(s_0 \ln(\Lambda/L_3)+\arctan(1/s_0))}~,
\label{eq:limcyc}
\eeq
where $c\approx 0.016$ is a normalization constant and
$s_0\approx 1.00624$ is a transcendental number that determines the
period of the limit cycle. If the cutoff $\Lambda$ is multiplied by
a factor $\exp(n\pi/s_0) \approx (22.7)^n$ with $n$ an integer, the
three-body coupling $\lambda_3$ is unchanged.
The three-body parameter $L_3$ can be determined directly from observable
quantities like the triton binding energy $B_t$.

Since all phenomenological nucleon-nucleon potentials are fitted to the same 
two-body data, they differ only in their off-shell properties.
It is well known, however, that short-distance three-body interactions and 
off-shell two-body interactions are related via unitary transformations 
(See, e.g., Ref.~\cite{Furnstahl:2000we} and references therein).
As a consequence, the difference between the various 
phenomenological nucleon-nucleon 
potentials corresponds at leading order in $\ell/|a|$ to
different values of the three-body parameter $L_3$.
The variation of $L_3$ parametrizes 
the Phillips and Tjon lines \cite{EfiTk85,Bedaque:1999ve,Platter:2004zs}.

Having expressed all relevant coupling constants 
in the effective interaction potential in terms of
physical observables, we are now ready to compute the triton
charge form factor.

\section{Triton Charge Form Factor}
\label{sec:emff}

At leading order (LO) in $\ell/|a|$, it is sufficient to apply the
electric charge operator to the wave function in the impulse
approximation. At next-to-leading order (NLO) the linear effective range
correction has to be included as well. For neutron-deuteron scattering,
various higher order calculations have been performed 
\cite{Hammer:2001gh,Bedaque:2002yg,Afnan:2003bs,Griesshammer:2004pe,
Griesshammer:2005ga}, but no calculations are available for
the triton charge form factor.
We note that using minimal substitution in the momentum dependent
potential (\ref{effpot})
will generate additional contributions from the regulator 
functions which raises the issue of gauge invariance.
However, as in the case of the momentum expansion of the effective 
potential discussed above these terms are 
at least of order  $k^2/\Lambda^2$. They can either be absorbed in
counterterms of the corresponding order (if present)
or shifted to higher orders 
by choosing regulator functions with sufficiently high powers of 
$n$ \cite{Epelbaum:1999dj}.
As a consequence, the effective theory calculation will be gauge 
invariant up to higher order corrections suppressed by  
$(k^2/\Lambda^2)^n$. In this paper, we focus on the universal 
aspects of the triton charge form factor and work at leading order.
Gauge dependent terms would enter at order $k^2/\Lambda^2$ which 
is beyond the accuracy of our calculation.
Higher order corrections will be considered explicitly 
in future work \cite{PlatteriP}.

A part of the following calculation has to be performed in
configuration space, thus
we define the corresponding Jacobi coordinates
${\bf x}= {\bf r}_1-{\bf r}_2$,
${\bf y}={\bf r_3}-{\textstyle\frac{1}{2}}({\bf r}_1+{\bf r}_2)$, and
${\bf R}={\textstyle\frac{1}{3}}({\bf r}_1+{\bf r}_2+{\bf r}_3)$.
The charge form factor of the triton is given by
\beq
\label{eq:ffdef}
F_C({\bf q}^2)=\langle\Psi_{{\bf K}+{\bf q}} \;{\bf k}_f|\rho_C|{\bf k}_i\;
\Psi_{\bf K}\rangle~,
\eeq
where ${\bf q}={\bf k}_i -{\bf k}_f$, ${\bf k}_i$ and ${\bf k}_f$ 
are the initial and final momentum of the scattered electron, and
$\Psi_{\bf K}$ denotes the full wave function with center of mass 
momentum ${\bf K}$.
The charge density operator $\rho_C$ is defined as \cite{Friar81}
\beq
\rho_C=\sum^3_i\Bigl[\frac{1}{2}(1+\tau_{iz})\rho_C^p({\bf r}-{\bf r}_i)
+\frac{1}{2}(1-\tau_{iz})\rho_C^n({\bf r}-{\bf r}_i)\Bigr]~,
\eeq
where $\rho_C^{n}$ and $\rho^p_C$ denote the charge densities of
the neutron and proton in configuration space, respectively,
and are related to the familiar Sachs form factors by
Fourier transformation.
The ${\bf r}_i$ denote the positions of the constituent nucleons. 
At LO the structure of the nucleons does not contribute and the
charge form factors of the proton and neutron are simply
$G_E^p=1$ and $G_E^n=0$. The first correction from nucleon
structure comes from the nucleon radii and would contribute
at N$^2$LO in our counting scheme. 

Due to the symmetry of the wave function, 
we can set $\rho_C=3 \rho_3$. Furthermore, we can 
drop constant overall factors as we will normalize $F_C$ at
${\bf q}^2=0$ to unity in the end.
Thus, by inserting a complete set of states into Eq.~(\ref{eq:ffdef})
and transforming to Jacobi coordinates ${\bf x}$, ${\bf y}$, and ${\bf R}$,
we obtain
\bea
\nonumber
F_C({\bf q}^2)&=&\int\hbox{d}^3 x\;\hbox{d}^3 y\;\hbox{d}^3 R
\sum_{i=1,2}\langle\Psi_{{\bf K}+{\bf q}}|{\bf x \;y \;R \;i}\rangle
\tilde\rho_3({\bf q})\exp(i \frac{2}{3} {\bf q \, y})\\
&&\qquad\qquad\qquad\qquad\qquad\qquad\qquad\times
\exp(i{\bf q \,R})
\langle {\bf x \;y \;R \;i}|\Psi_{\bf K}\rangle~.
\eea

Since the full wave function factorizes, 
$\langle \Psi_{\bf K}|{\bf x\, y\, R}\rangle =
\langle \Psi|{\bf x\, y}\rangle\cdot \exp(- i{\bf K R})$,
we can eliminate the center-of-mass integration.
Fourier transforming to momentum space, we finally obtain 
\bea
F_C({\bf q}^2)&=&\int\hbox{d}^3 u_1\;\hbox{d}^3 u_2\sum_{i=1,2}
\langle\Psi|{\bf u}_1 {\bf u}_2 {\bf i}\rangle \tilde\rho_3({\bf q})
\langle{\bf u}_1, {\bf u}_2 -\frac{2}{3}{\bf q},\; {\bf i}|\Psi\rangle~.
\eea
where ${\bf u}_1$ and ${\bf u}_2$ are the respective conjugate momenta.
Using Eq.~(\ref{eq:fullwf}), the form factor can be expressed in terms
of integrals over the first Faddeev component
$\psi_1$. The corresponding expressions are
somewhat lengthy and can be found in the Appendix.

\section{Results and Discussion}
\label{sec:resdis}

\begin{figure}[tb]
\centerline{\includegraphics*[width=12cm,angle=0]{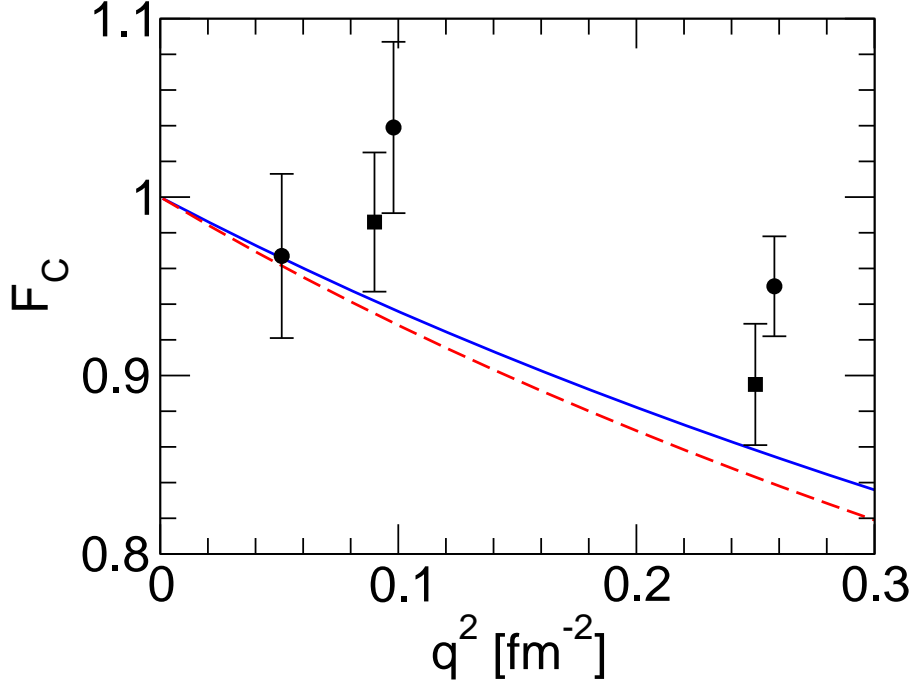}}
\caption{\label{fig:ff3H}The triton
charge form factor. The solid (dashed) line denotes the leading-order
result using $a_t$ and $a_s$ ($B_d$ and $a_s$) as two-body input.
The black circles and squares are experimental data
from Refs.~\cite{Beck84} and \cite{Beck87}, respectively.
}
\end{figure}

We are now in the position to calculate the charge form factor of
the triton. A simple estimate of higher order corrections can
be obtained by comparing the results from using the experimental
values for the triplet scattering 
length $a_t$ and the deuteron binding energy $B_d$ as two-body input.
To leading order in $\ell/|a_t|$, these two quantities are simply 
related via $B_d=1/a_t^2$; at higher orders they differ by effective
range terms. In Fig.~\ref{fig:ff3H}, we show the results
for the triton charge form factor at LO.
The solid line denotes the calculation
using $a_t$ and $a_s$ as two-body input, while 
the dashed line denotes the calculation using $B_d$ and $a_s$ as
two-body input. In both cases the three-body parameter $L_3$ is chosen
to reproduce the experimental value of the triton binding energy $B_t=8.48$ 
MeV. The corresponding values are $L_3 = 21.2$ fm$^{-1}$ for the solid line
and $L_3 = 19.8$ fm$^{-1}$ for the dashed line. The form factor results 
are plotted up to ${\bf q}^2=0.3$ fm$^{-2}$.
For momentum transfers of the order of the pion mass, 
the long-range character of one-pion exchange becomes important
and the pionless theory breaks down. 

The black circles and squares are experimental data
from Refs.~\cite{Beck84} and \cite{Beck87}, respectively.
While the general trend of the low-momentum transfer
form factor data is reproduced by our
calculation, we obtain a somewhat larger slope than the experiments.

The slope of the charge form factor at low momentum transfer
defines the triton charge radius:
\beq
\label{eq:chrad}
F_C({\bf q^2})= 1 - {\bf q^2} \langle r^2 \rangle/6 +\ldots\,.
\eeq
For notational simplicity, we use $r_C\equiv \langle r^2 \rangle^{1/2}$ 
in the following.

Fitting a polynomial in ${\bf q^2}$ to our result for the form factor, we
can extract the triton charge radius. We have fitted polynomials of 
varying degree up to 5th order in ${\bf q^2}$ and verified that
the extracted radius is independent of the degree of the polynomial.
This procedure leads to a triton charge radius $r_C=2.05$ fm 
for $a_t$ and $a_s$ as two-body input and  $r_C=2.18$ fm for
$B_d$ and $a_s$ as two-body input. The expected error of our LO 
calculation is $\ell/|a|\approx 30$\%. Averaging the results
from the two equivalent choices for the two-body input, we obtain  
$r_C^{LO}=(2.1\pm 0.6)$ fm. Our result is about 0.4 fm larger than the
experimental value $r_C^{exp}=(1.755\pm 0.086)$ fm \cite{Amroun94}
but both values are consistent within their error bars. It would 
be valuable to extend the calculation to higher orders
in order to judge the convergence of the expansion 
in $\ell/|a|$ \cite{PlatteriP}.
This requires including the linear effective range correction
at NLO. Higher order one-body terms from the nucleon radii contribute
only at N$^2$LO. Inclusion of the nucleon radii from 
Ref.~\cite{Hammer:2003ai} would increase
our result for $r_C$ by about 0.1 fm. The magnitude of this 
correction is consistent with our error estimate of 0.6 fm at LO.
We note, however, that there are additional contributions at N$^2$LO
such as a counter term that arises from gauging the momentum-dependent 
three-body force found in Ref.~\cite{Bedaque:2002yg}.

In Ref.~\cite{Friar85}, Friar, Gibson, Chen, and Payne observed 
a correlation between the triton charge radius and binding energy.
They explained this correlation as resulting from the primary
sensitivity of radii to the outer parts of the wave function
which are determined by the triton binding energy. Using the 
asymptotic form of the S-state wave function in hypersperical coordinates,
$\psi\propto\exp(-\rho \sqrt{B_t})/\rho^{5/2}$,
they were able to predict the leading functional dependence
\beq
r_C = b {B_t}^{-1/2}\,.
\label{eq:gibson}
\eeq
If the asymptotic form of the wave function is used for all
hyperradii $\rho$, one obtains $b=1/2$ \cite{Friar85}.
The correlation curve from Ref.~\cite{Friar85} is
reproduced by the phenomenological expression
\beq
\frac{r_C}{\rm fm} = 3.8 \left(\frac{B_t}{\rm MeV}\right)^{-0.41}\,.
\label{eq:gibsonfit}
\eeq

\begin{figure}[tb]
\centerline{\includegraphics*[width=11cm,angle=0]{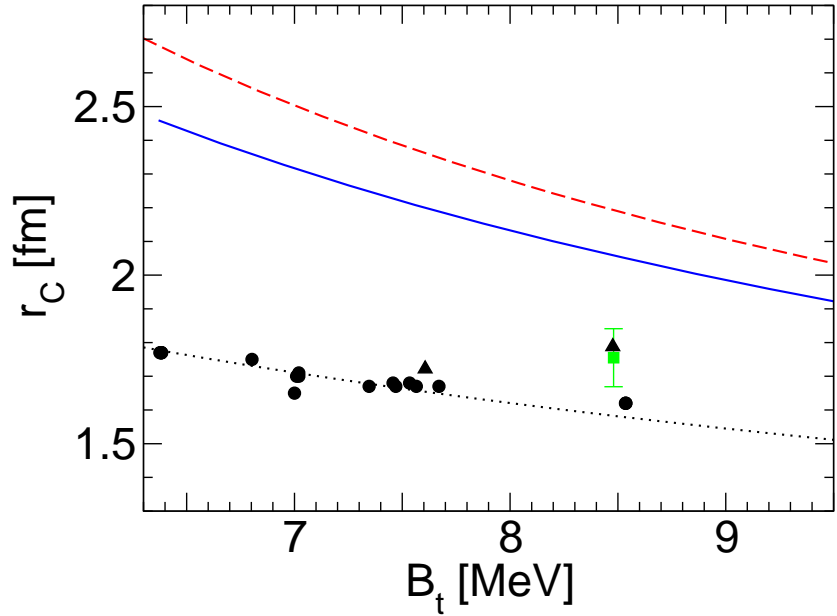}}
\caption{\label{fig:radius3H}
The correlation between the triton charge radius and binding energy. 
The solid (dashed) line denotes our leading-order
result using $a_t$ and $a_s$ ($B_d$ and $a_s$) as input parameters.
The circles indicate Faddeev calculations using different 
potentials from Refs.~\cite{Payne80,Chen85} while the two triangles show 
modern calculations using AV18 with and without the Urbana IX
three-body force \cite{Golak:2005iy}. The correlation curve of 
Ref.~\cite{Friar85} is given by the dotted line. The square 
indicates the experimental values.
}
\end{figure}

In Fig.~\ref{fig:radius3H}, we show the correlation 
between the triton charge radius and binding energy
at leading order in $\ell/|a|$.
The solid (dashed) line denotes our calculation
using $a_t$ and $a_s$ ($B_d$ and $a_s$) as input parameters.
Both lines are parametrized by the three-body parameter $L_3$. 
The circles indicate Faddeev calculations using different 
potentials from Ref.~\cite{Payne80,Chen85} while the correlation curve of 
Ref.~\cite{Friar85} is given by the dotted line.
The square gives the experimental values. 
The two triangles show modern calculations (including meson exchange
currents) based on the  AV18 potential with and without the Urbana IX
three-body force \cite{Golak:2005iy}. 
The potential models give the same two-body physics but
the corresponding values of $L_3$ are generally different.
As a consequence,
the observed correlation  between the triton charge radius and 
binding energy is naturally explained by variation of $L_3$. 
It has the same origin as the Phillips and Tjon lines.
We note that the dependence of the calculated correlation curves in 
Fig.~\ref{fig:radius3H} on the triton binding 
energy is similar to Eq.~(\ref{eq:gibson}) while the prefactor is different.
Using the same units as in Eq.~(\ref{eq:gibsonfit}),
we find a prefactor of 7.6 (9.5) and an exponent of $-0.61$ ($-0.69$)
for the solid (dashed) curves. This leads to a slightly stronger 
$B_t$ dependence in the leading order effective theory as in the
model calculations.

In the case of the Phillips line, the inclusion of the range 
correction moves the universal line closer to the experimental values
\cite{EfiTk85,Bedaque:2002yg}. Since the linear effective range correction
is the only new contribution at NLO, we expect a similar improvement
for the correlation between $r_C$ and $B_t$. A definite conclusion,
however, requires the explicit calculation of the range correction
which is beyond the scope of this study.

It is also interesting to note that there is no universal relation
between the binding energy and
the position of the minima and maxima of the form factor
\cite{Friar86} or the magnetic moments \cite{Tomu85}.
For the latter quantities, the universality is destroyed by meson exchange
currents. In the language of the pionless effective theory this 
corresponds to two-body currents entering at low orders in the expansion
in $\ell/|a|$. It would be very interesting to study the mechanism
that destroys universality for the magnetic moments. The position of
the  minima and maxima of the form factor (${\bf q }^2 \simgt 14$ fm$^{-2}$),
however, is outside the range of validity of the pionless effective theory.

\begin{acknowledgments}

We thank Ulf-G. Mei\ss ner  and A. Nogga  for discussions. 
HWH thanks the Institute for Nuclear Theory in Seattle where this work 
was  completed for its hospitality. This work was supported
in part by the EU Integrated Infrastructure Initiative Hadron Physics,
the Deutsche Forschungsgemeinschaft through funds provided 
to the SFB/TR 16 \lq\lq Subnuclear structure of matter'' and the
U.S. Department of Energy under grant DE-FG02-93ER40756.

\end{acknowledgments}

\appendix

\section{Faddeev Components}
\label{sec:fadd}
In the three-body system, it is always possible  to choose a cutoff 
$\Lambda$ for which the three-body force vanishes. 
Thus, the three-body force contribution can be dropped in the following.
Using Eq.~(\ref{eq:fullwf}), we express the full wave function $\Psi$ 
in terms of the first Faddeev component $\psi_1$ and work in a partial 
wave projected basis.
Defining $\hat{q}=|{\bf u}_2-{\textstyle\frac{2}{3}}{\bf q}|$, we have
\bea
\nonumber
F_C({\bf q}^2)&=&\int\hbox{d}u_1 u_1^2\,\hbox{d}u_2 u_2^2\int_{-1}^1\hbox{d}x
\Bigl\{\langle\Psi|u_1 u_2 {\bf 1}\rangle\tilde\rho_3(q)
\langle u_1 \hat q {\bf 1}|(1+P)|\psi_1\rangle\\
&&\qquad\qquad\qquad\qquad+
\langle\Psi|u_1 u_2 {\bf 2}\rangle\tilde\rho_3(q)
\langle u_1 \hat q \,{\bf 2}|(1+P)|\psi_1\rangle\Bigr\}~.
\eea
Applying the usual overlap equalities we obtain
\bea
\label{eq:fftriton}
\nonumber
F_C({\bf q}^2)&=&\int\hbox{d}u_1 u_1^2\,\hbox{d}u_2 u_2^2\int_{-1}^1\hbox{d}x
\Bigl\{\langle\psi_1|u_1 u_2\,{\bf 1}\rangle\tilde\rho_3(q)
\langle u_1 \hat q\, {\bf 1}|\psi_1\rangle
+
\langle\psi_1|u_1 u_2\,{\bf 2}\rangle\tilde\rho_3(q)
\langle u_1 \hat q\, {\bf 2}|\psi_1\rangle\\[0.3cm]
\nonumber
&&+
\int_{-1}^1\hbox{d}x'\Bigl[
\frac{1}{4}\langle\psi_1|\hat u_1(u1,u2)\,\hat u_2\,{\bf 1}
\rangle\tilde\rho_3(q)
\langle u_1 \hat q\, {\bf 1}|\psi_1\rangle
+
\frac{1}{4}\langle\psi_1|\hat u_1(u1,u2)\,\hat u_2\,{\bf 2}
\rangle\tilde\rho_3(q)
\langle u_1 \hat q\, {\bf 2}|\psi_1\rangle\\[0.3cm]
\nonumber
&&+
\frac{1}{4}\langle\psi_1|u_1 u_2\,{\bf 1}\rangle\tilde\rho_3(q)
\langle\hat u_1(u1,\hat q)\,\hat u_2\,{\bf 1}|\psi_1\rangle
+
\frac{1}{4}\langle\psi_1|u_1 u_2\,{\bf 2}\rangle\tilde\rho_3(q)
\langle\hat u_1(u1,\hat q)\,\hat u_2\,{\bf 2}|\psi_1\rangle\Bigr]\\[0.3cm]
\nonumber
&&+
\int_{-1}^1\hbox{d}x'\hbox{d}x''\Bigl[
\frac{5}{8}\langle\psi_1|\hat u_1(u1,u2)\,\hat u_2\,{\bf 1}
\rangle\tilde\rho_3(q)
\langle\hat u_1(u1,\hat q)\,\hat u_2\,{\bf 1}|\psi_1\rangle\\[0.3cm]
\nonumber
&&\qquad\qquad\qquad\qquad\qquad+
\frac{5}{8}\langle\psi_1|\hat u_1(u1,u2)\,\hat u_2\,{\bf 2}
\rangle\tilde\rho_3(q)
\langle\hat u_1(u1,\hat q)\,\hat u_2\,{\bf 2}|\psi_1\rangle
\Bigr]\Bigr\}~.
\eea
Now we separate the spin- and isospin components from the momentum space 
function
\bea
|{\bf 1}\rangle &=& \chi_1\eta_2~,\\
|{\bf 2}\rangle &=& \chi_2\eta_1~,
\eea
where $\chi$ denotes the spin- and $\eta$ the isospin part and
\bea
\nonumber
\langle u_1 u_2 {\bf 1}|\psi_1\rangle&\equiv&\langle u_1 u_2|\psi^{\bf 1}
\rangle \chi_1\eta_2~,\\
\langle u_1 u_2 {\bf 2}|\psi_1\rangle&\equiv&\langle u_1 u_2|\psi^{\bf 2}
\rangle \chi_2\eta_1~.
\eea
The isospin functions are written as
\bea
\eta_1&=&\frac{1}{\sqrt{6}}\Bigl[(--+)+(-+-)-2(+--)\Bigr]~,\\
\eta_2&=&\frac{1}{\sqrt{2}}\Bigl[(--+)-(-+-)\Bigr]~.
\eea
Thus, if we apply $\rho_3$ on the isospin functions we get
\bea
\tilde\rho_3(q)\eta_1&=&\frac{1}{\sqrt{6}}\Bigl[G_E^p(q)(--+)+G_E^n(q)(-+-)
-2G_E^n(q)(+--)\Bigr]~,\\
\tilde\rho_3(q)\eta_2&=&\frac{1}{\sqrt{2}}\Bigl[G_E^p(q)(--+)-G_E^n(q)(-+-)
\Bigr]~,
\eea
where $G_E^p$ and $G_E^n$ denote the proton and neutron charge form factors,
respectively.
Inserting the above definitions into Eq.~(\ref{eq:fftriton}) we obtain
\bea
\nonumber
\label{eq:formfactor2}
F_C({\bf q}^2) &=& G_E^p(q)\Bigl[
\frac{1}{2}I_1^{\bf 1}(q)+\frac{1}{6}I_1^{\bf 2}(q)+\frac{1}{8}I_2^{\bf 1}(q)
+\frac{1}{24}I_2^{\bf 2}(q)\\
\nonumber
&&\qquad\qquad\qquad\qquad+\frac{1}{8}I_3^{\bf 1}(q)+\frac{1}{24}I_3^{\bf 2}(q)
+\frac{5}{16}I_4^{\bf 1}(q)+\frac{5}{48}I_4^{\bf 2}(q)
\Bigr]
\\
&&+G_E^n(q)\Bigl[
\frac{1}{2}I_1^{\bf 1}(q)+\frac{5}{6}I_1^{\bf 2}(q)+\frac{1}{8}I_2^{\bf 1}(q)
+\frac{5}{24}I_2^{\bf 2}(q)\\
\nonumber
&&\qquad\qquad\qquad\qquad+\frac{1}{8}I_3^{\bf 1}(q)+\frac{5}{24}I_3^{\bf 2}(q)
+\frac{5}{16}I_4^{\bf 1}(q)+\frac{25}{48}I_4^{\bf 2}(q)
\Bigr]~,
\eea
where we have used the definitions
\bea
\nonumber
I_1^{\bf i} &=& \int\hbox{d}u_1 \,u_1^2\hbox{d}u_2 \,u_2^2\int_{-1}^1
\hbox{d}x\,\langle\psi_1^{\bf i}|u_1 u_1\rangle
\langle u_1 \hat q |\psi_1^{\bf i}\rangle~,\\
\nonumber
I_2^{\bf i} &=&  \int\hbox{d}u_1  \,u_1^2\hbox{d}u_2 \,u_2^2
\int_{-1}^1\hbox{d}x\int_{-1}^{1}\hbox{d}x'
\langle \psi_1^{\bf i}|\hat{u}_1(u_1,u_2)\hat{u}_2(u_1,u_2)\rangle
\langle u_1 \hat q|\psi_1^{\bf i}\rangle~,\\
\nonumber
I_3^{\bf i} &=& \int\hbox{d}u_1  \,u_1^2\hbox{d}u_2 \,u_2^2
\int_{-1}^1\hbox{d}x\int_{-1}^{1}\hbox{d}x'
\langle \psi_1^{\bf i}| u_1 u_2\rangle
\langle \hat{u}_1(u_1,\hat q)\hat{u}_2(u_1,\hat q)|\psi_1^{\bf 1}\rangle~,\\
\nonumber
I_4^{\bf i} &=& \int\hbox{d}u_1 \,u_1^2\hbox{d}u_2 \,u_2^2
\int_{-1}^1\hbox{d}x\int_{-1}^{1}\hbox{d}x'\int_{-1}^{1}\hbox{d}x'
\int_{-1}^1\hbox{d}x''
\langle \psi_1^{\bf i}|\hat{u}_1(u_1,u_2)\hat{u}_2(u_1,u_2)\rangle\\
&&\qquad\qquad\qquad\qquad\qquad\qquad\qquad\qquad\times
\langle \hat{u}_1(u_1,\hat q)\hat{u}_2(u_1,\hat q)|\psi_1^{\bf i}\rangle~.
\eea
At leading order, only the total charges 
of the proton and neutron contribute in Eq.~(\ref{eq:formfactor2}):
\beq
G_E^p ({\bf q}^2=0) = 1\qquad¸\mbox{and} \qquad
G_E^n ({\bf q}^2=0)=0 \,.
\eeq
The first correction from nucleon structure is due to the
charge radii of proton and neutron and enters at N$^2$LO.

\end{document}